\documentclass[aps,prd,twocolumn,a4paper,groupedaddress,amssymb,amsmath,nofootinbib,showpacs]{revtex4}
\usepackage[american]{babel}
\usepackage[T1]{fontenc}
\usepackage{psfrag,txfonts,graphicx,hyperref}

\DeclareMathOperator{\sgn}{sgn}				
\newcommand{\de}{\mathrm{d}}				
\newcommand{\pdiff}[2]{\frac{\partial\,#1}{\partial #2}}
\renewcommand{\vec}[1]{\ensuremath{\boldsymbol{#1}}}

\psfrag{H}{\(H\)}\psfrag{F}{\(F\)}\psfrag{D}{\(10(H-F)\)}\psfrag{G}{\(H=F\)}\psfrag{t}{\(t\)}\psfrag{P}{\(\phi\)}

\begin{document}

\title{Resolution of Curvature Singularities in Higher Derivative Gravity}
\preprint{LMU-ASC 13/08}

\author{Philipp \surname{H\"offer v.\ Loewenfeld}}
\email[]{phvl@theorie.physik.uni-muenchen.de}
\author{Ivo \surname{Sachs}}
\email[]{ivo@theorie.physik.uni-muenchen.de}
\affiliation{Arnold Sommerfeld Center for Theoretical Physics (ASC), Ludwig-Maximilians-Universit\"at M\"unchen, 80333 M\"unchen, Germany}
\date{June 20, 2008}

\begin{abstract}
We consider possible resolutions of singularities in a contracting anisotropic universe for a class of higher derivative gravity theories. We give 
evidence that for our models the big crunch singularity may be replaced by a nearly flat Minkowski-like phase before the Universe reenters an 
anisotropic expansion in a time-symmetric manner.
\end{abstract}

\maketitle

\section{Motivation and Results}

As a consequence of the singularity theorems of Penrose \cite{Penrose:1965} and Hawking 
\cite{Hawking:1976ra}, it is clear that Einstein's theory of gravity is incomplete. To be more specific: a spacetime obeying Einstein's field 
equations and some reasonable conditions on causality and energy contains incomplete, inextendable timelike or null geodesics and is thus singular in the 
Schmidt sense.

In general, the singularities predicted by these theorems are quite difficult to analyze, since the theorems do not predict their nature more concretely. But 
divergence of some curvature invariants seems to be a generic feature at any singularity in spacetime. In fact for singularities reached on timelike curves it 
can be proven (see \cite{Clarke:1975ph}) that the Riemann tensor becomes infinite. Thus effectively limiting the curvature seems to be a promising step 
on the way to a singularity free theory.

On the other hand, one expects that a more fundamental theory of gravity avoids singular solutions completely. Based on these observations we are led to 
construct approximations to the low energy effective theory of full quantum gravity as minimal deformations of  Einstein's general relativity by higher 
derivative terms in such a way that at least the curvature singularities are removed for generic solutions. A generic feature of low energy effective actions as 
limiting cases of fundamental theories of gravity is the appearance of higher derivative terms as well as nonlocal terms. At present we will not consider the 
latter. Similar higher derivative corrections as we use here are motivated from the low energy effective action of the massless boson sector of string theory in 
\cite{Easson:1999xw} and \cite{Easson:2003ia}. 

In general relativity there are infinitely many curvature invariants, e.g.
\[R, R_{\mu\nu}R^{\mu\nu}, R_{\mu\nu;\lambda}R^{\mu\nu;\lambda}, \ldots\]
which all have to be bounded in a nonsingular theory. The idea of the limiting curvature hypothesis is now to impose an explicit bound on a finite subset of 
invariants only and make sure, that their limiting value uniquely determines a nonsingular spacetime. In this way we ensure that any spacetime, that would be singular in Einstein gravity, is deformed into a nonsingular solution and henceforth all invariants are finite.

In practice, the limiting of a suitable invariant \(I\) can be achieved through the introduction of a Lagrange multiplier field \(\phi\) with an 
appropriate potential \(V(\phi)\). This idea was used by Brandenberger, Mukhanov, and Sornborger \cite{Brandenberger:1993ef} for 
homogeneous and isotropic spacetimes. 

In a Friedmann-Robertson-Walker cosmos with Hubble parameter \(H=\dot{a/a}\), the invariant 
\(I_2 = 4R_{\mu\nu}R^{\mu\nu}-R^2 = 12\dot{H}^2\) vanishes if and only if \(H=\text{const.}\) i.e.\ spacetime is de~Sitter. More generally, 
\(I_2=0\) characterizes de~Sitter spacetime as long as the Weyl tensor vanishes. Furthermore we remark that \(I_2\) is non-negative for a wide 
class of spacetimes (in particular, spherically symmetric ones).

The detailed analysis \cite{Brandenberger:1993ef} shows that the higher derivative theory with this invariant \(I_2\) has indeed no singular solutions, if we 
restrict ourselves to homogeneous and isotropic spacetimes. In this paper we investigate possible generalizations of this desingularization to anisotropic 
spacetimes. It is impossible to describe the corresponding dynamical system by analytical methods. For the model we consider in this work, we find evidence for 
the following generic behavior by a combination of analytical and numerical methods: (i) If the Lagrange multiplier field \(\phi\) grows large before the 
anisotropy becomes big then de~Sitter spacetime is an asymptotic attractor in the far future. (ii) If the initial anisotropies are small then the 
linearized evolution equations are compatible with the attractor solution. (iii) If the initial anisotropies are of order one, then a generic solution passes 
through a phase which is approximately Minkowski spacetime in finite time before returning to an anisotropic Universe in a time-symmetric fashion. This solution 
is nonanalytic. In particular, while all components of the Riemann tensor are finite, their derivatives are not. 

Apart from cosmological applications the higher derivative theory described here can also be applied to the interior of a black hole since the spacetime inside 
the horizon of a black hole resembles an anisotropic contracting, homogeneous universe. The mechanism suggested here would suffice to build nonsingular black 
holes. The surrounding of the singularity would then be replaced by a region which approaches flat Minkowski spacetime interpolating between a black 
hole and a white hole solution.

\section{The Model}
An invariant which measures the anisotropy of a spacetime is the squared Weyl tensor \(C^2 = C_{\mu\nu\tau\sigma}C^{\mu\nu\tau\sigma}\). To single out 
de~Sitter spacetime from the class of homogeneous and anisotropic spacetimes, we can thus choose the sum of \(C^2\), whose vanishing guarantees isotropy, 
and the non-negative invariant \(I_2\) selecting de~Sitter from the isotropic solutions. We thus choose the higher derivative action to be of the form
\begin{equation}\label{action}
  S = -\frac1{16\pi G}\int\left(R-I\phi+V(\phi)\right)\sqrt{-g}\de^4x.
\end{equation}
where the invariant \(I\) is given by
\begin{equation}
  I = I_2 + 3C^2 = 4R_{\mu\nu}R^{\mu\nu}-R^2+3C_{\mu\nu\tau\sigma}C^{\mu\nu\tau\sigma}.
\end{equation}
Here \(I_2\) is the invariant chosen in the isotropic case. For an isotropic Universe the invariant  \(I\) is, in fact, a perfect square. The factor 3 in front 
of the Weyl tensor squared is to simplify numerical factors, it does not alter the equations substantially, if one uses a different (but positive) factor. For 
the metric of an anisotropic and homogeneous cosmology (one special direction), we make the ansatz
\begin{equation}
  \de s^2 = -\nu(t)^2\de t^2+\alpha(t)^2\de x^2+\beta(t)^2\left(\de y^2+\de z^2\right),
\end{equation}
where the introduction of the lapse function \(\nu(t)\) is convenient to derive the field equations and set to \(1\) later. In this metric the invariant \(I\) 
takes the form
\begin{multline}
  I = 4\biggl(\left(\dot{F}+F^2-H^2\right)^2 +2\left(\dot{H}+H^2-HF\right)^2\\
      +\left(\dot{H}-\dot{F}-F^2+HF\right)^2\biggr).
\end{multline}
Here we introduced Hubble parameters \(H(t)=\frac{\dot{\beta}}{\beta}\) and \(F(t)=\frac{\dot{\alpha}}{\alpha}\) and set \(\nu(t)=1\) in the second line. 
Since \(I\) is the sum of three squares it vanishes if and only if all three terms vanish separately. There are two ways for this to happen. Either 
\begin{subequations}\begin{align}
  H(t) &= F(t) = \bar{H} = \text{const.}\label{I0dS},
\end{align}
which corresponds to the usual metric of spacially flat de~Sitter spacetime (or Minkowski spacetime for \(\bar{H}=0\)), or 
\begin{align}
  H(t) &= 0,\qquad F(t)=\frac1t\label{I0Mink},
\end{align}\end{subequations}
which describes Minkowski spacetime in unusual (singular) Kasner-like coordinates. 

Let us first discuss the conditions for approaching a de~Sitter Universe. If  \(V(\phi)\) rises slower than linear in \(\phi\) for \(\phi\to\infty\), the 
equations of motion derived from \eqref{action} will enforce de~Sitter spacetime as soon as \(\phi\) grows large. This is certainly achieved by a 
potential which approaches a constant value as \(\phi\) goes to infinity. 
On the other hand, to recover the usual Einstein theory of gravity as a low curvature limit, after elimination of the unphysical field \(\phi\) only 
higher order terms should remain. The auxiliary field is determined by the constraint equation
\begin{equation}\label{constraint}
  I = V'(\phi).
\end{equation}
Expanding \(V\propto\phi^m+\phi^{m+1}+\cdots\) for small \(\phi\) and keeping only the lowest order terms yields
\begin{equation}
  S \approx \begin{cases}\int\left(R + AI^{m/(m-1)}\right)\sqrt{-g}\de^4x &\text{for }m>1\\
                         \int\left(R + B + CI\right)\sqrt{-g}\de^4x &\text{for }m=1\end{cases}
\end{equation}
with some constants \(A, B, C\). Since \(I\) is itself quadratic in metric components, the correction term is of the desired order provided \(m>1\). The case 
\(m=1\) reduces to Einstein gravity with cosmological constant at low curvature. From the constraint equation \eqref{constraint} and the fact that the 
invariant \(I\) is strictly non-negative, it follows that \(V\) has to be a monotonically increasing function.

A potential satisfying the conditions given above is
\begin{equation}\label{potential}
  V(\phi) = H_0^2\frac{\phi^2\sgn(\phi)}{\left(1+\sqrt{\lvert\phi\rvert}\right)^4}
\end{equation}
More generally we may assume that the potential has an asymptotic power series in \(\frac1{\phi}\) of the form
\begin{equation}\label{app}
  V(\phi) = H_0^2\left(1-\frac{A}{\phi^n}+\frac{B}{\phi^{n+1}}+\ldots\right),
\end{equation}
where \(A\) is a positive constant and \(n>0\) (the special case \eqref{potential} corresponds to \(A=1\) and \(n=\frac12\)). We shall see below that the 
existence of an asymptotic de~Sitter attractor requires that \(n>1\). On the other hand one expects that the appearance of fractional powers in the 
asymptotic expansion \eqref{app} is a source for nonanalytic behavior. We will see below that this is indeed the case. 

\section{Field equations}
Variation of the higher derivative action \eqref{action} with respect to \ \(\phi\) gives (after setting \(\nu=1\))
\begin{subequations}\label{EOM}\begin{multline}\label{EOMphi}
  \left(\dot{F}+F^2-H^2\right)^2 + 2\left(\dot{H}+H^2-HF\right)^2 \\
   +\left(\dot{F}-\dot{H}+F^2-HF\right)^2 = \frac14V'(\phi), 
\end{multline}
while the 00-component of the Einstein equation takes the form
\begin{equation}\begin{split}\label{EOM00}
  0=\;&8\dot{\phi}\left((H+F)\dot{H}+H\dot{F}+2H(H-F)^2\right)\\
  +&4\phi\Bigl(2(H+F)\ddot{H} +2H\ddot{F} -\dot{H}^2 -2\dot{H}\dot{F} +4H(H+2F)\dot{H}\\
  &\qquad+2H(H+2F)\dot{F}-H^4+3H^2F^2-2HF^3\Bigr)\\
  -&2H^2-4HF-V,
\end{split}\end{equation}
where, again, we set \(\nu=1\) after variation.

The equations arising from variation w.r.t.\ the spatial components of the metric \(\alpha\) and \(\beta\) contain third derivatives of \(F\) and \(H\) but
given \eqref{EOMphi} and \eqref{EOM00} only one of the two spatial equations is independent. Thus we can choose a linear combination such that no third
derivatives of \(F\) appear (after setting \(\nu(t)=1\)):
\begin{equation}\begin{split}\label{EOMii}
  0=&8\ddot{\phi}\left(-\dot{H}+H(H-F)\right)\\
  +&4\dot{\phi}\Bigl(-4\ddot{H}-2\left(2H+3F\right)\dot{H}-2H\dot{F}\\
   &\qquad+4H^3-6FH^2+2F^2H\Bigr)\\
  +&4\phi\biggl(-2\dddot{H}-2\left(5H+F\right)\ddot{H}-7\dot{H}^2-2\dot{H}\dot{F}\\
   &\qquad-2\left(8H^2+4HF-F^2\right)\dot{H}-2\left(H^2-3HF\right)\dot{F}\\
   &\qquad-3H^4-2H^3F+5H^2F^2+2HF^3\biggr)\\
  +&4\dot{H}+6H^2+V.
\end{split}\end{equation}\end{subequations}

A closer look at the equations \eqref{EOM} reveals that the total differential order is \(5\). It is possible to eliminate the 
Lagrange multiplier field \(\phi\) with the help of \eqref{EOMphi}, which is algebraic in \(\phi\). The resulting system of 
two differential equations for \(\ddot{F}\) and \(\dddot{H}\) (or \(\dddot{F}\) and \(\ddot{H}\) ) contains the inverse function of the potential. Since we do 
not want to specify the potential beyond its asymptotic properties we choose to keep \(\phi\) in the equations. 

Let us now outline the strategy for solving the system of differential equations \eqref{EOM}. First we solve \eqref{EOMphi} for \(\dot{F}\). This involves 
taking the square root since \eqref{EOMphi} is a quadratic function in \(\ddot{F}\). We then use the resulting equation to eliminate  \(\dot{F}\) in 
\eqref{EOMphi} and solve for \(\dot{\phi}\). Using these equations we can eliminate all time derivatives of \(F\) and \(\phi\) in the spatial component of the 
field equation and thus obtain an equation for \(\dddot{H}\).

\section{Large \(\phi\) limit}
The case of large \(\phi\) bears special interest since this limit corresponds to the regime where the invariant has to be small. We will discuss the two 
configurations \eqref{I0dS} and  \eqref{I0Mink} with vanishing invariant \(I\) separately. In the first case we will assume that at a given time spacetime is 
approximately de~Sitter. Assuming the existence of a solution \(\phi(t)\) we can  replace the independent variable \(t\) by \(\phi\),  at least locally. 
The Hubble parameters and their derivatives can now be expanded in powers of \(\frac{1}{\phi}\). The \(\phi\)-equation \eqref{EOMphi} restricts the 
lowest order that can appear in these expansions, since each square ist at most of the order of \(\left(\frac{1}{\phi}\right)^{n+1}\):
\begin{align}\label{11}
  H &= \bar{H} + \frac{\bar{h}}{\phi^{(n+1)/4}} + \frac{h}{\phi^{(n+1)/2}} + \cdots,\\
  F &= \bar{H} + \frac{\bar{f}}{\phi^{(n+1)/4}} + \frac{f}{\phi^{(n+1)/2}} + \cdots,\\
  \dot{H} &= \frac{\tilde{h}}{\phi^{(n+1)/2}} + \cdots,\\
  \dot{F} &= \frac{\tilde{f}}{\phi^{(n+1)/2}} + \cdots
\end{align}
These expansions are not independent, since \(\dot{H}=\pdiff{H}{t}\), which will be used later to determine the coefficients \(\tilde{h}\) and \(\tilde{f}\). 
Substituting this expansion into \eqref{EOMphi} gives in leading order \(\bar{f}=\bar{h}\).\footnote{Actually selfconsistency later requires \(\bar{h}=0\).} The 
second derivatives of the Hubble parameters can be expressed through derivatives of \(\phi\):
\begin{subequations}\begin{align}
  \ddot{H} &= -\frac{\tilde{h}\frac{n+1}2}{\phi^{(n+1)/2}}\frac{\dot{\phi}}{\phi} + \cdots\\
  \ddot{F} &= -\frac{\tilde{f}\frac{n+1}2}{\phi^{(n+1)/2}}\frac{\dot{\phi}}{\phi} + \cdots
\end{align}\end{subequations}
With this expansion the \(00\)-equation \eqref{EOM00} reduces to:
\begin{equation}\begin{split}
  0= &4\dot{\phi}\left(\bar{H}\frac{n+3}2(\tilde{f}+2\tilde{h})\frac1{\phi^{(n+1)/2}}+\mathcal{O}\left(\frac1{\phi^{3(n+1)/4}}\right)\right)\\
  +&4\phi\left(3\bar{H}^2(\tilde{f}+2\tilde{h})\frac1{\phi^{(n+1)/2}}+\mathcal{O}\left(\frac1{\phi^{3(n+1)/4}}\right)\right)\\
  -&3\bar{H}^2 - \frac{H_0^2}2 + \mathcal{O}\left(\frac1{\phi^{(n+1)/4}}\right).
\end{split}\end{equation}
Neglecting higher order terms we get
\begin{equation}\label{largephilimit}
  \dot{\phi}+\bar{H}\frac32(n+3)\phi-\frac{\frac34\bar{H}+\frac38\frac{H_0^2}{\bar{H}}}{\tilde{f}+2\tilde{h}}\phi^{(n+1)/2}=0.
\end{equation}
For \(n<1\) a collapsing universe (\(\bar{H}<0\)) then leads to an asymptotically exponential growth
\begin{equation}
  \phi(t) \propto \exp\left(-\frac{3(n+3)}2\bar{H}t\right),
\end{equation}
so that \(\phi\) increases forever. For \(n=1\) there is a second term of order \(\phi\) in \eqref{largephilimit}, which leads to an asymptotic behavior of 
the form
\begin{equation}
  \phi(t) \approx \exp\left(\left(\frac{3\bar{H}^2+\frac12H_0^2}{8\bar{H}(\tilde{f}+2\tilde{h})}-\frac32\bar{H}\right)t\right).
\end{equation}
This means for \(n=1\) \(\phi=\infty\) (i.e.\ de~Sitter spacetime) is an attractor only if \(\tilde{f}+2\tilde{h}<0\) or 
\(\tilde{f}+2\tilde{h}>\frac14+\frac1{24}\frac{H_0^2}{\bar{H}}\). These conditions can however not be assumed \emph{a priori}. They can only be verified once the 
asymptotic solution is known. This is in contrast to the isotopic model \cite{Brandenberger:1993ef} where exponential growth of \(\phi\) is generic.

With the help of the asymptotic equation for \(\dot{\phi}\)
\begin{equation}
  \dot{\phi} = -3\bar{H}\phi + \mathcal{O}\left(\phi^{3/4}\right)
\end{equation}
we can check the consistency of our ansatz \eqref{11} for \(H\) and \(\dot{H}\) (and analogously for \(F\) and \(\dot{F}\)), since
\begin{equation}
  \dot{H} = \pdiff{H}{\phi}\dot{\phi} = -\frac{\frac{n+1}4\bar{h}}{\phi^{(n+1)/4}}\frac{\dot{\phi}}{\phi} - \frac{\frac{n+1}2h}{\phi^{(n+1)/2}}\frac{\dot{\phi}}{\phi} + \mathcal{O}\left(\frac1{\phi^{3(n+1)/4}}\right)
\end{equation}
By comparing coefficients we get
\begin{subequations}\begin{align}
  \bar{h}&=0 & \tilde{h}&=\frac32(n+1)\bar{H}h\\
  \bar{f}&=0 & \tilde{f}&=\frac32(n+1)\bar{H}f
\end{align}\end{subequations}
therefore the leading terms in the expansion of \(H,F,\dot H,\dot F\) in \eqref{11} are all determined in terms of \(h\) and \(f\). 

\begin{figure}
\includegraphics[width=\linewidth]{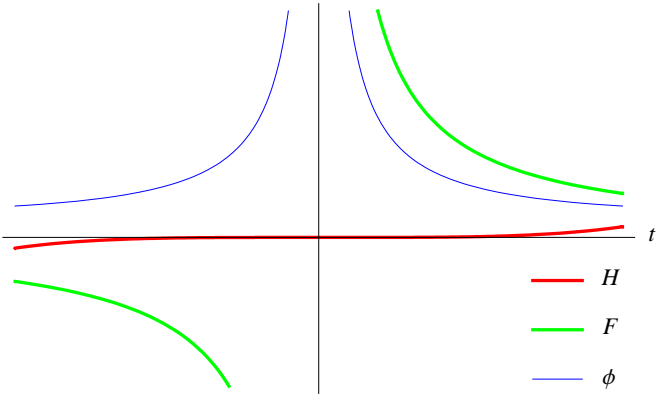}
\caption{Qualitative behavior of a possible solution bouncing through Minkowski} spacetime. \(\phi(t)\) diverges symmetrically around \(t=0\) while  \(F(t)\)  diverges symmetrically. \(H(t)\) is smooth at \(t=0\). 
\label{singsketch}
\end{figure}

Let us now turn to the second possibility \eqref{I0Mink} for vanishing invariant \(I\) at large \(\phi\). Here the Hubble parameters approach 
Minkowski spacetime in singular coordinates \eqref{I0dS}, i.e.
\begin{equation}\label{Minksing}
  H(t) \to 0, \qquad F(t)^2 + \dot{F}(t) \to 0, \qquad t\to t_0.
\end{equation}
Without restricting the generality we can assume \(t_0=0\). In order to see if this behavior can be realized in a global solution we are then looking for a 
solution of the field equations approaching \(H=0\) and \(F=\frac1t\) for \(t\to0\). In this case the field equations immediately imply \(\phi\to\frac1t\). 
To continue we observe that the field equations are invariant under the discrete transformation simultaneously sending \(t\to-t\), \(H(t)\to-H(t)\), 
\(F(t)\to-F(t)\), and 
\(\phi(t)\to\phi(t)\). Approaching \(t_0=0\) from below with these asymptotics, we thus suggest the behavior sketched in Fig.~\ref{singsketch}.

To make this discussion more precise let us rewrite the equations of motion in terms of \(G=1/F\), \(H\), and \(\psi=1/\phi\) which are more suitable for analyzing the small \(t\) behavior.
The proposed asymptotic solution is now \(G=t\), \(\psi=\lvert{}t\rvert\), and \(H=0\). Therefore we seek approximate solutions of the form
\begin{align}\label{generpower}
 G &= t +a_1t^{1 +\alpha_1} +a_2t^{1 +\alpha_1 +\alpha_2} +a_3t^{1 +\alpha_1 +\alpha_2 +\alpha_3} +\ldots, \\
 H &= b_1t^{\beta_1} +b_2t^{\beta_1 +\beta_2} +b_3t^{\beta_1 +\beta_2 +\beta_3} +\ldots, \\
 \psi &= \pm t \pm c_1t^{1 +\gamma_1} \pm c_2t^{1 +\gamma_1 +\gamma_2} \pm c_3t^{1 +\gamma_1 +\gamma_2 +\gamma_3} +\ldots,
\end{align}
where \(\alpha_i\), \(\beta_i\), and \(\gamma_i\) are positive constants and the two signs correspond to \(t\gtrless0\). In the following we will only discuss 
the case \(t>0\). For convenience we choose a specific potential with asymptotic expansion
\begin{equation}
 V(\phi) = V_1 -\frac{4V_2}{\phi^{1/2}} +\frac{10V_3}{\phi} -\frac{20V_4}{\phi^{3/2}} +\frac{35V_5}{\phi^2} \mp\ldots
\end{equation}
The potential \(V(\phi)=H_0^2\frac{\phi^2}{\left(1+\sqrt{\phi}\right)^4}\) corresponds to the choice \(V_i=H_0^2\). Plugging this ansatz into the 
equations of motion and neglecting all obviously subleading terms we remain with
\begin{subequations}\begin{align}
 0 &= -\frac{W}{4}t^{3/2} +b_1^2(3 -2\beta_1 +3\beta_1^2)t^{2(\beta_1 -1)}\label{LO1} \\
   &\qquad+2a_1b_1(1 +\alpha_1)(1 +\beta_1)t^{-3 +\alpha_1 +\beta_1} +2a_1^2(1 +\alpha_1)^2t^{2\alpha_1 -4},\notag\\
 0 &= V_1 +8b_1(2 +\beta_1 -\beta_1^2)t^{\beta_1 -4},\label{LO2} \\
 0 &= V_1 -8b_1(6 +\beta_1 -4\beta_1^2 +\beta_1^3)t^{\beta_1 -4}.\label{LO3}
\end{align}\end{subequations}
In \eqref{LO2} and \eqref{LO3} we can either choose \(\beta_1=4\) and cancel both terms against each other or we have to choose \(\beta_1<4\) as a zero of the 
two coefficients. We are left with two admissible solutions obeying \(\beta_1>0\):
\begin{subequations}\begin{align}
 \alpha_1 &= \frac{11}4, & \beta_1 &= 2, & a_1 &= \pm\frac{\sqrt{V_2}}{13},&\qquad\text{or }\\
 \alpha_1 &= \frac{11}4, & \beta_1 &= 4, & a_1 &= \pm\frac{\sqrt{V_2}}{13}, & b_1 &= \frac{V_1}{80}.
\end{align}\end{subequations}
Proceeding in this way we obtain a perturbative solution such that the equations of motion are satisfied up to linear order at \(t=0\) with
\begin{subequations}\label{singsol}\begin{align}
 G &= t -\frac{2V_2}{77b_1}t^{7/2}, \\
 H &= b_1t^2 +(48b_1c_1 +V_1)t^4 +\frac{2(1155b_1c_2 -38V_2)}{4235}t^{9/2}, \\
 \psi &= \pm t \pm c_1t^3 \pm c_2t^{\frac72}\qquad\text{for } t\gtrless0.
\end{align}\end{subequations}
We have not been able to extend the solution as a power series in fractional powers of \(t\) beyond this order suggesting that, if a solution exists for a 
finite range of \(t\), nonanalytic behavior of a different kind will be required. Note also that all components of the Riemann tensor (but not all its 
derivatives) are finite at \(t=0\).

To summarize, we found that if \(\phi\) is already large while the two Hubble parameters are comparable the proposed limiting curvature procedure works 
perfectly well. In the case where the leading order correction to the potential has an exponent \(n<1\), large \(\phi\) implies that spacetime is nearly 
de~Sitter and the evolution of \(\phi\) extends to infinite future while growing exponentially and thus forcing spacetime to approach de~Sitter 
even more. Because of the exponential growth of \(\phi\) it takes an infinite time to actually reach the de~Sitter end stage. On the other hand, if the 
difference of the two Hubble parameters is of order one while \(\phi\) is  large, the metric may approach Minkowski spacetime in singular 
coordinates in finite time. Although this solution, if it exists, is nonanalytic at \(t=0\), it can be continued symmetrically through the ``Minkowski 
phase'' at \(t=0\). In Sec.~\ref{numsol} we will further analyze the existence of a global solution using numerical methods.

\section{Phase Space Analysis}
In this section we discuss the global properties of the solutions of the equation of motion by means of a phase space analysis. The equations of motion 
\eqref{EOM} form a system of ordinary differential equations with total differential order of 5. We eliminate derivatives of \(F\) from the equations 
\eqref{EOM00} and \eqref{EOMii} with the help of \eqref{EOMphi} and all derivatives of \(\phi\) from Eq.~\eqref{EOMii} by \eqref{EOM00}. As usual we 
rewrite the equations into a system of first order differential equations for the vector \(\vec{u}\coloneqq\left(\phi,F,H,\dot{H},\ddot{H}\right)\) and obtain
\begin{equation}
 \dot{\vec{u}} = \vec{V}(\vec{u})
\end{equation}
Since \eqref{EOMphi} contains \(\dot{F}^2\) solving the equations involves a square root of
\begin{equation}
 B = V'(\phi)-10F^2H^2 + 20FH^3-10H^4+12H(F-H)\dot{H}-10\dot{H}^2.
\end{equation}
The reality of the solution is thus not guaranteed \emph{a priori}. However, it turns out that the hypersurface \(B=0\) is left invariant under the flow of the vectorfield \(\vec{V}\), since
\begin{equation}
 \left.\pdiff{B}{u^i}V^i\right|_{B=0}=0.
\end{equation}
Hence, if we restrict the initial conditions \(u^i_{(0)}\) to the domain of \(B>0\), i.e.\ real vector field \(\vec{V}(\vec{u}_{(0)})\), the solution will remain 
real for all times.

Let us now discuss the existence of singular points. Critical points where \(\vec{V}(\vec{u})\) vanishes are problematic if they correspond to fixed points away 
from the asymptotic isotropic regime. Since \(V^3=u^4\) and \(V^4=u^5\), we can reduce the problem of searching for critical points \(\vec{V}=\vec{0}\) to two 
dimensions by replacing \(u^4\) and \(u^5\) with 0, and \(u^2\) by one of the four solutions of the equation \(V^2=0\). It can be shown that the remaining two 
functions \(V^1(u^1,u^2)\) and  \(V^5(u^1,u^2)\) do not vanish simultaneously. But \(V^5(u^1,u^2)\) has a one-dimensional set of fractional zeros and a one-
dimensional set of fractional poles. The poles typically correspond to branching points of the solution. If more than one real solution meet at a branching 
point, then the Cauchy problem is not well defined. If furthermore the set of poles and zeros intersect then further complications arise since the 
details will now depend on how the solution approaches the singularity. We can not exclude the existence of such points; but a numerical investigation of \(\left\lvert V^5(u^1,u^2)\right\rvert\) indicates that whether the intersection is empty or not depends on the details of the potential \(V(\phi)\).

\section{Numerical solutions}\label{numsol}
\begin{figure}
 \includegraphics[width=\linewidth]{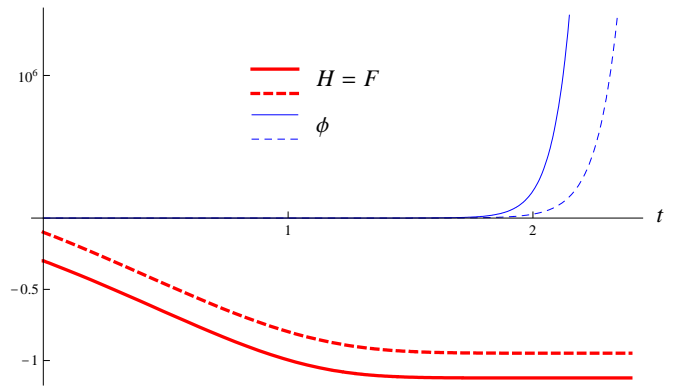}
 \caption{Two numerical solutions (solid and dashed lines; \(H=F\) (thick, negative axis), \(\phi\) (thin, positive axis)) starting with isotropic initial conditions.}
 \label{num_isotropic}
\end{figure}
In this section we supplement the analytic discussion of the set of solutions to the higher derivative action by a numerical investigation. Of particular 
interest is the question of whether the nonanalytic local solution passing through the Minkowski stage can be embedded in a global solution and to 
determine the fate of a generic anisotropic Universe. While we are not able to give a conclusive answer to these questions, we find that the numerical analysis 
supports the idea that a generic anisotropic initial condition will go through a Minkowski phase before returning to the anisotropic Universe in a time-symmetric 
fashion. 

Let us begin however with isotropic initial conditions, the behavior of the solutions of \cite{Brandenberger:1993ef} is recovered --- the Hubble 
parameter \(H(t)=F(t)\) approaches a constant while \(\phi\) grows exponentially. The two numerical solutions plotted in Fig.~\ref{num_isotropic} were 
obtained using initial conditions
\begin{align*}
 H(0)=F(0)&=\begin{cases}-0.1\\-0.3\end{cases}, & \phi(0)&=0.1.
\end{align*}
For these and further numerical solutions we chose the potential to be of the form \eqref{potential} with the constant set to \(H_0=10\).

\begin{figure}
 \includegraphics[width=\linewidth]{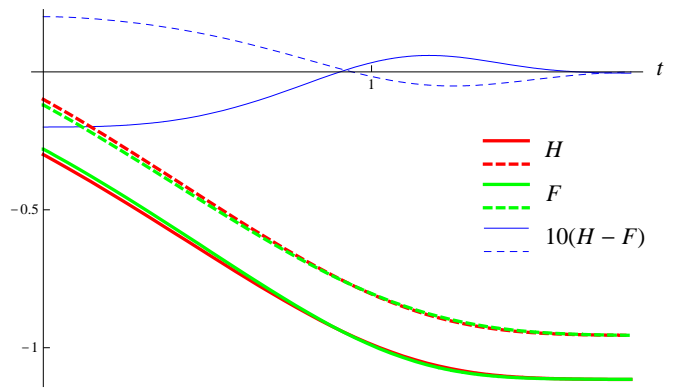}
 \caption{Two numerical solutions (solid and dashed lines; \(H\), \(F\) (thick), and \(H-F\) (thin) where the latter is rescaled with a factor of \(10\)) starting  with initial conditions slightly perturbed from isotropy.}
 \label{num_pertisotropic}
\end{figure}
Initial conditions prescribing a slightly perturbed isotropic universe evolve to an isotropic solution --- the perturbations are damped and the Hubble 
parameters and \(\phi\) show the same asymptotic behavior as in the isotropic case. Fig.~\ref{num_pertisotropic} shows the difference \(H-F\) decay while 
the two Hubble parameters approach a constant. The unphysical field \(\phi\) is not plotted since its behavior is similar to the isotropic case. The
initial conditions used were
\begin{align*}
 H(0)&=\begin{cases}-0.1\\-0.3\end{cases}, & F(0)&=\begin{cases}-0.12\\-0.28\end{cases}, & \phi(0)&=0.1.
\end{align*}

This numerical prediction can be supported by considering the linearization of the system of equations around an isotropic background with \(F(t)=H(t)\) 
satisfying the 00-component of the field equations with \(\phi\) given by the constraint equation, we obtain a system of quasilinear equations. The linearized 
system of field equations decouples after introduction of the new dependent variables \(x(t)=\delta F(t) - \delta H(t)\) and 
\(y(t)=\delta F(t) + 2\delta H(t)\). The equation for the linearized anisotropy \(x(t)\) is
\begin{multline}
 0 = \dddot{x} +2\left(3H +\frac{\dot{\phi}}{\phi}\right)\ddot{x} +\left(15H^2 +6\dot{H} +3H\frac{\dot{\phi}}{\phi} +\frac{\ddot{\phi}}{\phi} -\frac1{2\phi}\right)\dot{x}\\
 +3\left(6H^3+7\dot{H}H+\ddot{H}\right)x.
\end{multline}
Hence, a hint on the linear stability of the isotropic solution can be derived from the eigenvalues of this differential equation. With the help of the 
background field equations, we can express the eigenvalues in terms of \(H\) and \(\phi\). In a contracting universe (\(H<0\)) the real part of at least one 
eigenvalue is positive if \(\phi\) and/or \(H\) are small which might result in an instability. Hence, we conclude that the isotropic solution is stable if 
\(\phi\) is large. Furthermore, a numerical analysis shows that the linear perturbations stay small long enough for the background field \(\phi\) to become 
large. Note that, since the square of the Weyl tensor vanishes in the isotropic background together with its first derivative, there are no contributions 
from the Weyl tensor in the linearized equations. Thus, they are no longer applicable, as soon as the background field \(\phi\) becomes large and thus the 
nonlinear constraint equation \eqref{EOMphi} becomes important.

The numerical evolution of generic initial conditions breaks down at a pole like singularity in \(F\), while \(H\) approaches \(0\), and \(\phi\) has just 
passed a minimum and thus seems to be approximately constant. But from \eqref{EOMphi} we imply that \(\phi\propto\frac1t\) as soon as \(F\) approaches 
the singularity further. This behavior resembles the aforementioned singular coordinate system of Minkowski spacetime. We therefore suggest that these 
numerical solutions could be patched to the approximate solution \eqref{singsol}  describing a kind of bounce through flat 
Minkowski spacetime.

\begin{figure}
 \includegraphics[width=\linewidth]{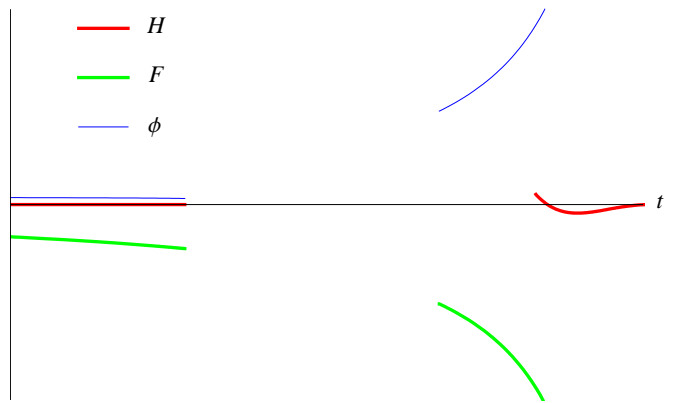}
 \caption{Sketch of the proposed matching of the numerical solution on the left and the possible approximate continuation on the right. There is no overlap between the respective regions of validity of the two approximations.}
 \label{numcont}
\end{figure}
In Fig.~\ref{numcont} this continuation is sketched. The numerical solution is integrated from the left until it stops. Using the free parameters that are 
left in \eqref{singsol} the asymptotical solution is fitted approximately. We do not expect to be able to patch the two sets of functions together 
smoothly since there is no overlap between the range of validity of the numerical solution and the approximate solution \eqref{singsol}  at $t=t_0$.
In particular \(H\) obviously does not meet. Nevertheless the picture hints on the existence of a continued solution.

\section{Conclusions}
In this paper we considered a higher derivative theory of gravity in which the square of the Weyl tensor together with another invariant, \(I_2\), whose 
vanishing guarantees de~Sitter spacetime, is bounded with the help of a Lagrange multiplier field. For a generic model of this type we find two 
qualitatively different end points of a contracting, anisotropic Universe: If the initial anisotropies are small then the anisotropies stay small in the 
linearized regime for long enough allowing the Lagrange multiplier field to grow. We found convincing evidence that for this set of initial configurations the 
nonlinear effects will then suppress the anisotropies leading to an asymptotic contracting de~Sitter spacetime for late times. Since the 
Weyl tensor appears only quadratically in the higher derivative theory the anisotropies are not suppressed in the linear regime. A possible way to improve on 
this in the linear regime might be to introduce separate Lagrange multipliers  for the Weyl tensor and \(I_2\), respectively. This would probably 
help to enlarge the regions of applicability of the linearized theory. 

On the other hand, if the initial anisotropies are of order 1, then the linearized theory does not apply. A combination of analytical methods and numerical 
approximations in this regime produced circumstantial evidence that a global solution may exist in this case which interpolates between a contracting 
anisotropic Universe and a nearly Minkowski phase in finite time before returning to an anisotropic spacetime in a time-symmetric fashion. While we are 
not able to prove the existence of such a solution globally at present, we think that a better understanding of this solution, if it exists, could have important 
applications in the resolution of cosmological singularities as an alternative to the so-called bounce solutions. Furthermore the same mechanism could be used 
to resolve the spacelike singularity inside a Schwarzschild black hole.
In our model we find evidence that the surrounding of the singularity is replaced by a region which approaches flat Minkowski spacetime interpolating 
between a black hole and a white hole solution. This resembles the bridge between two large classical solutions obtained from a quantum geometrical discussion in \cite{Ashtekar:2005qt}. In an alternative scenario proposed in \cite{Frolov:1988vj}, the black hole singularity is replaced by a region 
which is approximately de~Sitter spacetime. While we do not find evidence for this scenario in our model we can not exclude it at present.

\begin{acknowledgments}
We would like to thank V. Mukhanov for initial collaboration on the project and for helpful discussions throughout the completion of the paper. The work was 
supported in parts by the Transregio TRR~33 ``The Dark Universe'' and the Excellence Cluster ``Origin and Structure of the Universe'' of the DFG. P.\;H.\;v.\;L.\ is supported 
by the IMPRS on Elementary Particle Physics of the MPI for Physics.
\end{acknowledgments}


\begin{thebibliography}{8}
\expandafter\ifx\csname natexlab\endcsname\relax\def\natexlab#1{#1}\fi
\expandafter\ifx\csname bibnamefont\endcsname\relax
  \def\bibnamefont#1{#1}\fi
\expandafter\ifx\csname bibfnamefont\endcsname\relax
  \def\bibfnamefont#1{#1}\fi
\expandafter\ifx\csname citenamefont\endcsname\relax
  \def\citenamefont#1{#1}\fi
\expandafter\ifx\csname url\endcsname\relax
  \def\url#1{\texttt{#1}}\fi
\expandafter\ifx\csname urlprefix\endcsname\relax\def\urlprefix{URL }\fi
\providecommand{\bibinfo}[2]{#2}
\providecommand{\eprint}[2][]{\url{#2}}

\bibitem[{\citenamefont{Penrose}(1965)}]{Penrose:1965}
\bibinfo{author}{\bibfnamefont{R.}~\bibnamefont{Penrose}},
  \bibinfo{journal}{Phys. Rev. Lett.} \textbf{\bibinfo{volume}{14}},
  \bibinfo{pages}{57} (\bibinfo{year}{1965}).

\bibitem[{\citenamefont{Hawking}(1976)}]{Hawking:1976ra}
\bibinfo{author}{\bibfnamefont{S.~W.} \bibnamefont{Hawking}},
  \bibinfo{journal}{Phys. Rev.} \textbf{\bibinfo{volume}{D14}},
  \bibinfo{pages}{2460} (\bibinfo{year}{1976}).

\bibitem[{\citenamefont{Clarke}(1975)}]{Clarke:1975ph}
\bibinfo{author}{\bibfnamefont{C.~J.~S.} \bibnamefont{Clarke}},
  \bibinfo{journal}{Commun. Math. Phys.} \textbf{\bibinfo{volume}{41}},
  \bibinfo{pages}{65} (\bibinfo{year}{1975}).

\bibitem[{\citenamefont{Easson and Brandenberger}(1999)}]{Easson:1999xw}
\bibinfo{author}{\bibfnamefont{D.~A.} \bibnamefont{Easson}} \bibnamefont{and}
  \bibinfo{author}{\bibfnamefont{R.~H.} \bibnamefont{Brandenberger}},
  \bibinfo{journal}{JHEP} \textbf{\bibinfo{volume}{09}}, \bibinfo{pages}{003}
  (\bibinfo{year}{1999}), \eprint{arXiv:hep-th/9905175}.

\bibitem[{\citenamefont{Easson}(2003)}]{Easson:2003ia}
\bibinfo{author}{\bibfnamefont{D.~A.} \bibnamefont{Easson}},
  \bibinfo{journal}{Phys. Rev.} \textbf{\bibinfo{volume}{D68}},
  \bibinfo{pages}{043514} (\bibinfo{year}{2003}),
  \eprint{arXiv:hep-th/0304168}.

\bibitem[{\citenamefont{Brandenberger et~al.}(1993)\citenamefont{Brandenberger,
  Mukhanov, and Sornborger}}]{Brandenberger:1993ef}
\bibinfo{author}{\bibfnamefont{R.~H.} \bibnamefont{Brandenberger}},
  \bibinfo{author}{\bibfnamefont{V.~F.} \bibnamefont{Mukhanov}},
  \bibnamefont{and}
  \bibinfo{author}{\bibfnamefont{A.}~\bibnamefont{Sornborger}},
  \bibinfo{journal}{Phys. Rev.} \textbf{\bibinfo{volume}{D48}},
  \bibinfo{pages}{1629} (\bibinfo{year}{1993}), \eprint{arXiv:gr-qc/9303001}.

\bibitem[{\citenamefont{Ashtekar and Bojowald}(2006)}]{Ashtekar:2005qt}
\bibinfo{author}{\bibfnamefont{A.}~\bibnamefont{Ashtekar}} \bibnamefont{and}
  \bibinfo{author}{\bibfnamefont{M.}~\bibnamefont{Bojowald}},
  \bibinfo{journal}{Class. Quant. Grav.} \textbf{\bibinfo{volume}{23}},
  \bibinfo{pages}{391} (\bibinfo{year}{2006}), \eprint{arXiv:gr-qc/0509075}.

\bibitem[{\citenamefont{Frolov et~al.}(1990)\citenamefont{Frolov, Markov, and
  Mukhanov}}]{Frolov:1988vj}
\bibinfo{author}{\bibfnamefont{V.~P.} \bibnamefont{Frolov}},
  \bibinfo{author}{\bibfnamefont{M.~A.} \bibnamefont{Markov}},
  \bibnamefont{and} \bibinfo{author}{\bibfnamefont{V.~F.}
  \bibnamefont{Mukhanov}}, \bibinfo{journal}{Phys. Rev.}
  \textbf{\bibinfo{volume}{D41}}, \bibinfo{pages}{383} (\bibinfo{year}{1990}).

\end{thebibliography}

\end{document}